\documentclass[sigconf]{acmart}
\usepackage{tabularx}
\usepackage{hyperref}
\usepackage{cleveref}
\usepackage{diagbox}
\usepackage{multirow}
\usepackage{caption}
\usepackage{graphicx}
\usepackage{float} 
\usepackage{amsmath}  %
\usepackage{balance}
\usepackage{subcaption}
\usepackage{makecell}
\usepackage[most]{tcolorbox}
\crefname{figure}{Fig.}{Fig}
\crefname{table}{Table}{Table}
\crefname{equation}{Eq.}{eq}
\crefname{section}{Section}{Section}
\usepackage{enumitem}  

\newcolumntype{L}[1]{>{\raggedright\arraybackslash}p{#1}}
 
\newcolumntype{C}[1]{>{\centering\arraybackslash}p{#1}}
 
\newcolumntype{R}[1]{>{\raggedleft\arraybackslash}p{#1}}
\AtBeginDocument{%
  }

\setcopyright{acmlicensed}
\copyrightyear{2018}
\acmYear{2018}
\acmDOI{XXXXXXX.XXXXXXX}
\acmConference[Conference acronym 'XX]{Make sure to enter the correct
  conference title from your rights confirmation email}{June 03--05,
  2018}{Woodstock, NY}
\acmISBN{978-1-4503-XXXX-X/2018/06}

\begin{document}

\title{SynH-Rank: Quality-Aware Code Search via Diverse Data Synthesis and Hierarchical Ranking Training}

\author{Keyu Liang}
\orcid{0009-0000-4613-247X}
\affiliation{%
  \department{College of Computer Science and Technology and The State Key Laboratory of Blockchain and Data Security}
  \institution{Zhejiang University}
  \city{Hangzhou}
  \country{China}
}
\email{liangkeyu@zju.edu.cn}

\author{Haoye Wang}
\orcid{0000-0002-3314-0427}
\affiliation{%
  \institution{Hangzhou City University}
  \city{Hangzhou}
  \country{China}
}
\email{wanghaoye@hzcu.edu.cn}

\author{Yanfu Yan}
\orcid{0009-0008-2475-6802}
\affiliation{%
  \institution{Zhejiang University and Hangzhou High-Tech Zone (Binjiang) Institute of
  Blockchain and Data Security}
  \city{Hangzhou}
  \country{China}
}

\email{yanfu@zju.edu.cn}

\author{Zhiyuan Wan}
\orcid{0000-0001-7657-6653}
\affiliation{%
  \department{College of Computer Science and Technology and The State Key Laboratory of Blockchain and Data Security}
  \institution{Zhejiang University}
  \city{Hangzhou}
  \country{China}
}
\email{wanzhiyuan@zju.edu.cn}

\author{Zhongxin Liu}
\orcid{0000-0002-1981-1626}
\authornote{Zhongxin Liu is the corresponding author.}
\affiliation{%
  \department{College of Computer Science and Technology and The State Key Laboratory of Blockchain and Data Security}
  \institution{Zhejiang University}
  \city{Hangzhou}
  \country{China}
}
\email{liu_zx@zju.edu.cn}

\renewcommand\footnotetextcopyrightpermission[1]{}
\settopmatter{printacmref=false} 
\renewcommand{\shortauthors}{Trovato et al.}

\begin{abstract}
Code search is crucial for developer productivity because it enables the efficient reuse of existing code.
Modern code search systems typically adopt a retrieve-then-rerank pipeline, where rerankers refine the initial results by modeling semantic relevance between queries and candidates. 
However, these rerankers focus solely on semantic relevance and overlook critical non-functional qualities such as execution speed, memory usage, and maintainability. 
In practice, such qualities are far from negligible: studies show that many developers expect search results to maintain a high standard of coding style and have specific needs such as optimizing code for resource usage. 
This highlights the need for integrating quality awareness into code search for practical software development. 

Achieving quality-aware code search is challenging for two main reasons. 
First, quality-annotated datasets are scarce, making effective training challenging.
Second, standard contrastive learning objectives are not well-suited to quality-aware ranking.
Contrastive learning has become the dominant training approach for rerankers because it is highly effective at distinguishing relevant candidates from irrelevant ones. 
However, its binary objective of optimizing for relevance versus irrelevance struggles to capture the ordinal relationships among high-quality relevant code, low-quality relevant code, and irrelevant code.
To address these challenges, we propose SynH-Rank, a quality-aware code reranking framework that combines LLM-driven diverse data synthesis with hierarchical ranking training.
SynH-Rank utilizes a three-level labeling scheme to explicitly model the ranking hierarchy: high-quality relevant $>$ low-quality relevant $>$ irrelevant. 
We further introduce a new benchmark containing 4,209 pairs, along with two novel metrics: Quality Preference Accuracy (QPA), which measures the model's ability to prioritize high-quality code, and Multi-Condition Accuracy (MCA), which evaluates performance in complex multi-constraint scenarios. 
Experimental results show SynH-Rank improves QPA by 20.15\% over backbone models and outperforms standard relevance-only contrastive training by 15.80\%, while simultaneously enhancing traditional relevance metrics and multi-condition generalizability.
\end{abstract}



\received{20 February 2007}
\received[revised]{12 March 2009}
\received[accepted]{5 June 2009}

\maketitle

\newtcolorbox{highlightbox}{%
  breakable,
  enhanced,
  colback=blue!5!gray!10,   %
  colframe=blue!40!black,   %
  coltitle=black,           %
  boxrule=0.8pt,            %
  arc=2mm,                  %
  left=6pt,right=6pt,top=6pt,bottom=6pt,
  before skip=8pt,after skip=8pt,
}

\section{Introduction}

Code search aims to retrieve code snippets that are semantically relevant to a given natural language query.
As a fundamental task in software engineering, code search significantly enhances developer productivity by facilitating code reuse~\cite{di2023code}.
More recently, it has become a critical component in retrieval-augmented generation (RAG) and in-context learning frameworks, further highlighting its importance~\cite{zhang2025coderag}.

Despite its success, existing code search research primarily focuses on semantic relevance, often overlooking the non-functional quality of retrieved code.
In practice, developers have requirements well beyond functional correctness: studies show that 73.58\% of developers expect search results to maintain a high standard of coding style, while many also prefer results that are concise and representative~\cite{liu2024empirical}.
Furthermore, \cite{singhal2024nofuneval} emphasizes that functional correctness alone is insufficient in practical software development; developers must also consider quality-related requirements, such as optimizing Java code for resource usage on memory-constrained Android devices.
Consequently, incorporating quality awareness into code search is essential, as it profoundly impacts both developer experience and the broader software engineering lifecycle.

While significant efforts have been made to assess and improve the quality of code generation~\cite{singhal2024nofuneval,guo2024codeeditorbench,peng2025perfcodegen}, research considering code quality within the context of code search remains limited.
Recent studies have started to focus on code quality evaluation, finding that current retrieval models struggle to distinguish between high-quality and low-quality code~\cite{geng2025coquir}. 
However, these studies primarily focus on benchmark construction and problem diagnosis, leaving the question of how to train models to handle these challenges largely unexplored.
Moreover, existing benchmarks~\cite{geng2025coquir,singhal2024nofuneval} evaluate quality preferences along isolated attributes, failing to capture real-world scenarios where developers impose multiple simultaneous constraints (e.g., “find code that is both fast and memory-efficient to implement functionality X”).

Modern code search systems commonly adopt a retrieve-then-rerank pipeline~\cite{hu2023revisiting, multi}, where a reranker refines an initial set of candidates. This architecture offers strong potential to enforce quality-aware constraints by incorporating fine-grained quality preferences during the reranking stage.
Rerankers are capable of performing detailed relevance assessment by jointly modeling the query and candidate code, enabling them to capture nuanced quality signals (e.g., efficiency, readability, or memory usage). In particular, the deep interaction architecture of rerankers (e.g., cross-encoders)~\cite{gotmare2021cascaded} provides sufficient capacity to model subtle semantic and structural cues that simpler retrievers may overlook.
Moreover, rerankers can serve as plug-and-play components that are seamlessly integrated into existing pipelines, avoiding the substantial cost of re-indexing while still enabling flexible and effective quality-aware refinement.

Despite this potential, traditional reranking methods primarily focus on relevance optimization, and extending them to a quality-aware paradigm introduces several non-trivial challenges: 
(1) Data scarcity. 
Training data with explicit contrastive quality labels is limited~\cite{du2024mercury}. 
Moreover, gathering such data is challenging because measuring dynamic properties like memory usage is a complex process~\cite{blanco2022software}, especially when code snippets are incomplete and cannot be executed.
(2) Neglect of fine-grained ranking order. Existing reranking methods typically employ the InfoNCE (Information Noise-Contrastive Estimation) loss~\cite{oord2018representation} to distinguish relevant code from irrelevant negatives. However, this binary formulation overlooks the inherent ranking structure among candidates, such as: high-quality relevant code > low-quality relevant code > irrelevant code. Treating low-quality yet relevant code as standard negative samples not only fails to reflect developer quality preferences but may also weaken the model’s ability to preserve correct relevance judgments.

To address these challenges, we first construct a benchmark derived from real-world code data encompassing three critical quality dimensions: execution speed, memory usage, and maintainability. These three dimensions strike a balance between expressiveness and practicality: compared to binary relevance labels, they enable a more fine-grained characterization of developer preferences, while avoiding the increased annotation cost and complexity associated with introducing a larger number of dimensions. This benchmark comprises 4,209 pairs and is designed to evaluate both quality preference and traditional relevance in code reranking. We further extend it to multi-condition scenarios to assess model generalizability under complex constraints. To quantify these capabilities, we introduce two metrics: Quality Preference Accuracy (QPA) and Multi-Condition Accuracy (MCA).

Subsequently, we propose \textbf{SynH-Rank}, a framework for quality-aware code search based on diverse data \textbf{Syn}thesis and \textbf{H}ierarchical ranking training. 
Specifically, we leverage Large Language Models (LLMs) with specialized guidelines to synthesize diverse high-quality and low-quality code variants for a given query. 
To incorporate quality awareness into the supervision, we adopt a three-level labeling scheme that categorizes candidates into high-quality relevant, low-quality relevant, and irrelevant code. 
We then formulate the training process as a learning-to-rank task, employing a hierarchical ranking loss to explicitly capture the preference order: high-quality relevant $>$ low-quality relevant $>$ irrelevant. 
This approach enables SynH-Rank to go beyond binary relevance and effectively discriminate fine-grained quality signals during reranking.

Experimental results on our Python benchmark covering execution speed, memory usage, and maintainability demonstrate that state-of-the-art rerankers, such as Qwen3-Reranker-0.6B, still lack sufficient quality-aware capabilities in these settings. 
Compared with the backbone models, SynH-Rank improves the QPA metric by an average of 20.15\% across the three evaluated models. 
Furthermore, SynH-Rank outperforms standard relevance-only contrastive learning by 15.80\% in QPA. Notably, our framework also improves traditional relevance metrics. 
For instance, compared with standard relevance-only contrastive training, it improves MRR@10 by 4.64\% on Jina-multilingual-reranker-v2-base and by 2.49\% on Qwen3-Reranker-0.6B. 
Additional analyses show that SynH-Rank generalizes effectively to multi-conditional reranking scenarios within the evaluated quality dimensions, establishing a strong baseline for Python quality-aware code search. The main contributions of this paper are as follows:
\begin{itemize}[leftmargin=1.2em]
    \item We construct a quality-aware code reranking benchmark covering three quality dimensions, namely execution speed, memory usage, and maintainability. The benchmark contains 4,209 pairs and is accompanied by two new metrics, QPA and MCA, for evaluating quality preference and multi-condition generalization.

    \item We propose \textbf{SynH-Rank}, a framework that combines LLM-based diverse data synthesis with hierarchical ranking training over a three-level hierarchy, thereby explicitly integrating quality-awareness into reranking supervision.

    \item Extensive experiments show that SynH-Rank improves QPA by 20.15\% over backbone models and outperforms standard relevance-only contrastive training by 15.80\%, while also improving traditional relevance metrics and generalizing to multi-condition scenarios.
\end{itemize}

\section{Background and Related Work}
\subsection{Background}
Code search is a fundamental task in software engineering that aims to retrieve the most relevant code snippets from a large-scale repository in response to a natural language query. 
Modern code search systems typically adopt a retrieve-and-rerank (or recall-and-rerank) architecture to balance efficiency and effectiveness~\cite{hu2023revisiting}.
The \textit{retrieval} stage first identifies a small set of candidates by calculating the similarity between the query and code, which are typically encoded into separate vector embeddings. 
In the \textit{reranking} stage, the query is concatenated with each candidate code snippet individually to compute a more precise relevance score. 
This joint processing enables a deeper semantic interaction than independent encoding.

\subsection{Code Search}
\subsubsection{Retriever}
Modern code retrievers~\cite{guo2022unixcoder,guo2020graphcodebert,shi2023cocosoda,feng2020codebert,chen2025hedgecode,liang2025zero} typically adopt a bi-encoder~\cite{karpukhin2020dense} architecture to support efficient similarity search in high-dimensional vector spaces, with a primary focus on learning effective code representations~\cite{fan2025exploring,zhang2023vulnerability,liu2024pre,yan2024enhancing}.
CodeBERT~\cite{feng2020codebert} established the foundation by leveraging masked language modeling for cross-modal alignment.
Building upon this, UniXcoder~\cite{guo2022unixcoder} utilizes a unified framework with multimodal constraints to enhance semantic representations.
Recent advanced methods like HedgeCode~\cite{chen2025hedgecode} improves retrieval by introducing a multi-task hedging contrastive learning framework to capture fine-grained semantic distinctions.

\subsubsection{Reranker}
Most rerankers adopt a \textbf{cross-encoder} architecture based on BERT-like models.
Given a query $q$ and a code snippet $d$, the model concatenates them into the sequence ``\texttt{[CLS] $q$ [SEP] $d$ [SEP]}'', feeds it into the encoder, and passes the \texttt{[CLS]} representation to a linear layer to produce a relevance score.
Currently, BGE-Reranker~\cite{xiao2024c} has become one of the most widely adop\-ted rerankers due to its strong performance benefiting from large-scale multi-task pre-training.
Jina Reranker~\cite{jina-reranker-v2} further improves reranking quality on code search tasks through multilingual training and carefully designed hard negative mining, showing advantages over BGE-based models.

More recently, \textbf{generative rerankers}\cite{nogueira2020document,reddy2024first,zhang2025qwen3} have emerged as a competitive alternative by leveraging the reasoning capacity of large language models.
Instead of encoding query-document pairs, these methods formulate relevance estimation as a conditional generation problem.
Qwen3-Reranker~\cite{zhang2025qwen3}, for instance, adopts a decoder-only architecture and estimates relevance as:
\begin{equation}
  s(q, d) = \frac{e^{P(\text{yes} \mid \mathcal{I}, q, d)}}{e^{P(\text{yes} \mid \mathcal{I}, q, d)} + e^{P(\text{no} \mid \mathcal{I}, q, d)}}
\end{equation}
where $\mathcal{I}$ is a task-specific instruction prompt. 
In this work, we investigate both cross-encoder-based and generative rerankers to improve  the performance of quality-aware code search.

\subsection{Quality-aware Software Engineering}
Ensuring code quality is a critical objective in software engineering~\cite{tornhill2022code}. 
With the growing use of LLMs in software development~\cite{shao2025unigencoder}, the community has increasingly recognized that merely satisfying functional requirements is insufficient.
Retrieved or generated code must also adhere to high-quality standards in order to be robust and safe for deployment in production environments.

Code generation research has increasingly focused on quality-aware modeling and evaluation~\cite{liu2024no,singhal2024nofuneval,du2024mercury}. Beyond traditional metrics like Pass@k, recent works have introduced fine-grained evaluation frameworks. NoFunEval~\cite{singhal2024nofuneval} evaluates whether code LLMs can satisfy non-functional requirements such as efficiency and maintainability, finding that many models struggle with these constraints even when generating functionally correct code. EffiLearner~\cite{huang2024effilearner} leverages execution feedback to guide LLMs in self-optimization for code efficiency. SmellCC~\cite{xue2025clean} reconstructs training data to improve the maintainability of LLM-generated code.
These methods are complementary to our work but target a different setting: they improve or evaluate the code that a generator produces, whereas a search reranker must choose among pre-existing candidates and cannot assume permission to edit them.
This difference makes the ranking objective central, because the model must preserve functional relevance while expressing a preference among candidates with different quality levels.

Despite these advances in generation, quality-aware research in the context of code search remains limited. 
Most existing code search benchmarks, such as CodeSearchNet~\cite{husain2019codesearchnet}, primarily focus on semantic relevance between a query and a code snippet. 
Recent studies~\cite{geng2025coquir} have indicated that retrieval models struggle to differentiate code quality, yet these works focus primarily on problem diagnosis rather than providing optimization frameworks.
Moreover, existing benchmarks typically treat quality attributes in isolation, and therefore fail to evaluate scenarios with multiple simultaneous constraints (e.g., ``find code that is both fast and
memory-efficient to implement functionality X''). 
Addressing these limitations via \textit{reranking} offers a practical advantage, as it can be integrated into existing retrieve-and-rerank pipelines without requiring the re-indexing of massive code repositories. 
This highlights the need for specialized training frameworks for quality-aware ranking, as well as benchmarks capable of evaluating such multi-condition environments.

\section{Dataset Construction}
Existing benchmarks largely overlook code quality and multi-cond\-ition scenarios. 
To address this, we construct QualCode, a benchmark covering three quality dimensions: speed, memory, and maintainability, following NoFunEval~\cite{singhal2024nofuneval}. 
\textbf{QualCode} supports both quality preference evaluation and relevance evaluation. 
We further extend this to a multi-condition setting, proposing \textbf{MC-QualCode} to evaluate model performance on queries with multiple constraints.

Inspired by CoQuIR~\cite{geng2025coquir}, we build our benchmark on top of CodeNet~\cite{puri2021codenet}, a large-scale dataset of code samples collected by IBM from various online judge websites.
CodeNet contains submission metadata such as acceptance status, CPU time, and memory usage.
Among the 4,053 problems in CodeNet, we exclude those with non-English problem descriptions. 
To ensure that our benchmark focuses on quality rather than functional correctness, we retain only the submissions marked as ``accepted'', so that the selected answers correspond correctly to the problem statements.
\cref{tab:benchmark} summarizes the key statistics of the datasets.

\subsection{Information Retrieval Dataset: QualCode}\label{sec:dataset_ir}
For each problem, we construct a set of $\langle i, q, p, n \rangle$ tuples, 
where $i$ denotes a quality-dimension instruction (e.g., \textit{``Retrieve code that is fast/memory-efficient/maintainable''}), $q$ is the problem description, $p$ is a 
high-quality solution, and $n$ is a low-quality solution with respect to the 
specified dimension.
This dataset supports two tasks:

\textbf{Information Retrieval Task}: Given $\langle i, q \rangle$, the model retrieves code from a corpus, where $p$ serves as the ground truth positive, and all solutions to other problems are treated as negatives. 

\textbf{Quality Preference Task}: Given a tuple $\langle i, q, p, n \rangle$, the model is expected to assign a higher score to $p$ than to $n$ under the quality constraint specified by $i$.

The construction process for each dimension is as follows:

\textbf{Speed}.  
We identify the code with the minimum CPU time as the positive example and the one with the maximum CPU time as the negative example. 
To ensure a clear distinction in performance levels between the positive and negative examples, we impose the following condition:  $2 \times \text{cpu\_time}_{positive} < \text{cpu\_time}_{negative}$.

\textbf{Memory}.
The construction process for the memory dimension follows the same procedure as that for the speed dimension, except that memory usage is used as the selection criterion.

\textbf{Maintainability}.
To evaluate code maintainability, we utilize SonarQube, a widely recognized static analysis tool, to quantify code smells as a key metric.
We randomly select a code sample with no code smells as the positive example. 
The negative example is chosen as the code that contains the highest number and diversity of code smells.

\begin{figure*}[!t]
    \centering
    \includegraphics[width=\textwidth]{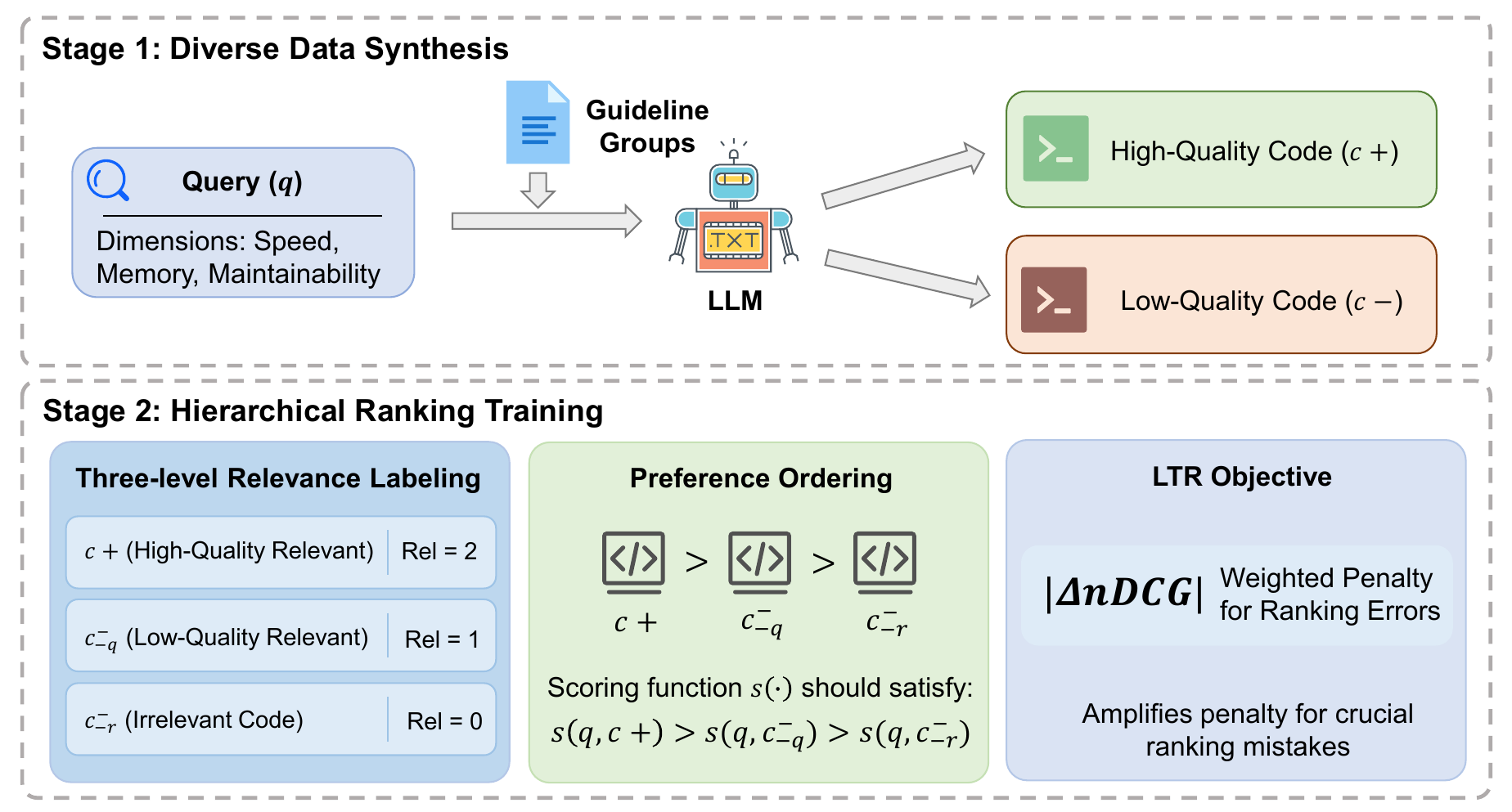}
    \caption{Overall Framework of SynH-Rank.}
    \label{fig:overview}
\end{figure*}
\label{sec:method}

\subsection{Multi-condition Dataset: MC-QualCode}\label{sec:dataset_multi-con}
In real-world software development, user queries often require results that satisfy multiple constraints simultaneously, such as being fast and memory-efficient. 
To reflect this need in code search, we construct a multi-condition ranking dataset to evaluate whether models can prioritize code that meets composite instructions. 
Each sample is represented as a triplet $\langle i, q, p, n \rangle$, where:
\begin{itemize}[leftmargin=1.2em]
    \item $i$: An instruction specifying the desired conditions, e.g., \textit{``Retrieve code that is fast and memory-efficient."}
    \item $q$: The problem description.
    \item $p$: A positive sample satisfying all $N$ conditions specified in $q$.
    \item $n$: A hard negative satisfying $N-1$ conditions. 
    Following~\cite{lu2025multiconir},  we use samples satisfying $N-1$ conditions as hard negatives, since distinguishing one missing condition is more challenging than distinguishing larger gaps. 
    This choice also implements a controlled-variable strategy: the positive and negative candidates differ in only one requested quality dimension, which isolates the model's sensitivity to the missing condition.
\end{itemize}

Since we have three quality dimensions, we create three sub-datasets covering 1-condition, 2-condition, and 3-condition tasks to evaluate model performance across varying levels of constraint complexity, each containing 3,000 randomly sampled triplets.
For each condition level, we include all possible combinations of the specified quality dimensions.
We sample examples in a balanced manner across combinations within each subset to avoid bias toward any particular quality requirement.
The samples are derived from the CodeNet corpus, with binary quality labels assigned along three dimensions:
\begin{itemize}[leftmargin=1.2em]
    \item \textbf{Speed \& Memory}: 
    Solutions in the top 33\% (lowest CPU time and memory usage) are labeled 1, and those in the bottom 33\% are labeled 0. 
    The middle tier is discarded to ensure a clear quality boundary between positive and negative samples.
    \item \textbf{Maintainability}: Samples without code smells are labeled 1, while those with code smells are labeled as 0.
\end{itemize}

Based on these labels, we apply a controlled-variable strategy to construct triplets.
For a $k$-condition instruction, $p$ satisfies all $k$ conditions specified in $i$, while $n$ satisfies exactly $k-1$ conditions.

\subsection{Evaluation Method}\label{sec:dataset_eval}
We evaluate code search performance from two perspectives: \textbf{relevance} and \textbf{quality preference}.

\textbf{Relevance}. Given a query and a set of code snippets, the goal of a code search method is to retrieve relevant code snippets.
To evaluate the relevance of retrieved code snippets, we adopt two widely used metrics: Normalized Discounted Cumulative Gain (\textbf{nDCG}) and Mean Reciprocal Rank (\textbf{MRR}). These metrics quantify the performance of the code search system in ranking relevant code snippets higher.
In our setup, high-quality relevant code snippets are treated as positive examples. To reduce computational overhead during evaluation, we randomly sample 99 irrelevant code snippets as negative examples for each query.

\textbf{Quality preference}. Given a query and a set of relevant code snippets, the goal of the code search method is to prioritize high-quality code snippets. 
To measure this ability, we introduce a metric called Quality Preference Accuracy (QPA), defined as:
\begin{equation}
QPA = \frac{1}{|S|} \sum_{\langle i,q, p, n \rangle \in S} \mathbb{I}[\text{score}(i, q, p) > \text{score}(i,q, n)]
\end{equation}

Here, $|S|$ is the total number of triplets, $ \mathbb{I}[\cdot] $ is the indicator function, $i$ denotes the instruction, $ q $ represents the query, $ p$ denotes the positive (high-quality) code snippet, $ n$ denotes the negative (low-quality) code snippet, and $\text{score}(\cdot, \cdot)$ denotes the scoring function of the code search method.
QPA captures the proportion of triplets for which the method correctly ranks the positive snippet higher than the negative snippet for a given query.

\textbf{Multi-condition Evaluation}. To assess the model's ability to handle complex composite constraints, we evaluate performance on the 1-condition, 2-condition, and 3-condition sub-datasets. 
For these tasks, we utilize \textbf{Multi-Condition Accuracy (MCA)} as an evaluation metric. Similar to QPA, MCA measures the frequency with which the model ranks a sample satisfying all $N$ conditions ($p$) higher than a hard negative sample satisfying only $N-1$ conditions ($n$). 
This evaluation strictly tests the model's sensitivity to fine-grained constraint satisfaction.

\begin{table}[!htbp]
  \centering
    \belowrulesep=0pt
  \aboverulesep=0pt
  \caption{Detailed Benchmark Statistics}
  \label{tab:benchmark}
  \small %
  \renewcommand\arraystretch{1.3} %
  \setlength\tabcolsep{2pt} %
  
  \begin{tabularx}{8cm}{C{2.5cm}|C{1.2cm}|C{2.5cm}|C{1.2cm}}
    \toprule
    \textbf{QualCode} & \textbf{\# Pair} & \textbf{MC-QualCode} & \textbf{\# Pair} \\
    \midrule
    Speed           & 1524 & 1-cond. & 3000 \\
    Memory          & 1429 & 2-cond. & 3000 \\
    Maintainability & 1256 & 3-cond. & 3000 \\
    \bottomrule
  \end{tabularx}
\end{table}

\section{Approach}

Figure~\ref{fig:overview} illustrates the overall architecture of \textbf{SynH-Rank}.
The framework consists of two core components: \textit{diverse data synthesis} (\S\ref{sec:synthesis}), which generates quality-aware training data through guideline-driven LLM prompting, and \textit{hierarchical ranking training} (\S\ref{sec:train}), which learns the fine-grained preference hierarchy among candidates of varying quality and relevance.
In this section, we describe each component in turn.

\subsection{Diverse Data Synthesis}
\label{sec:synthesis}

A fundamental bottleneck in quality-aware code search is the scarcity of labeled data covering non-functional properties.
To overcome this, we synthesize paired code samples that differ in quality across three non-functional dimensions: \textit{speed}, \textit{memory}, and \textit{maintainability}.

\begin{figure}[!htbp]
  \centering

  \includegraphics[width=\columnwidth]{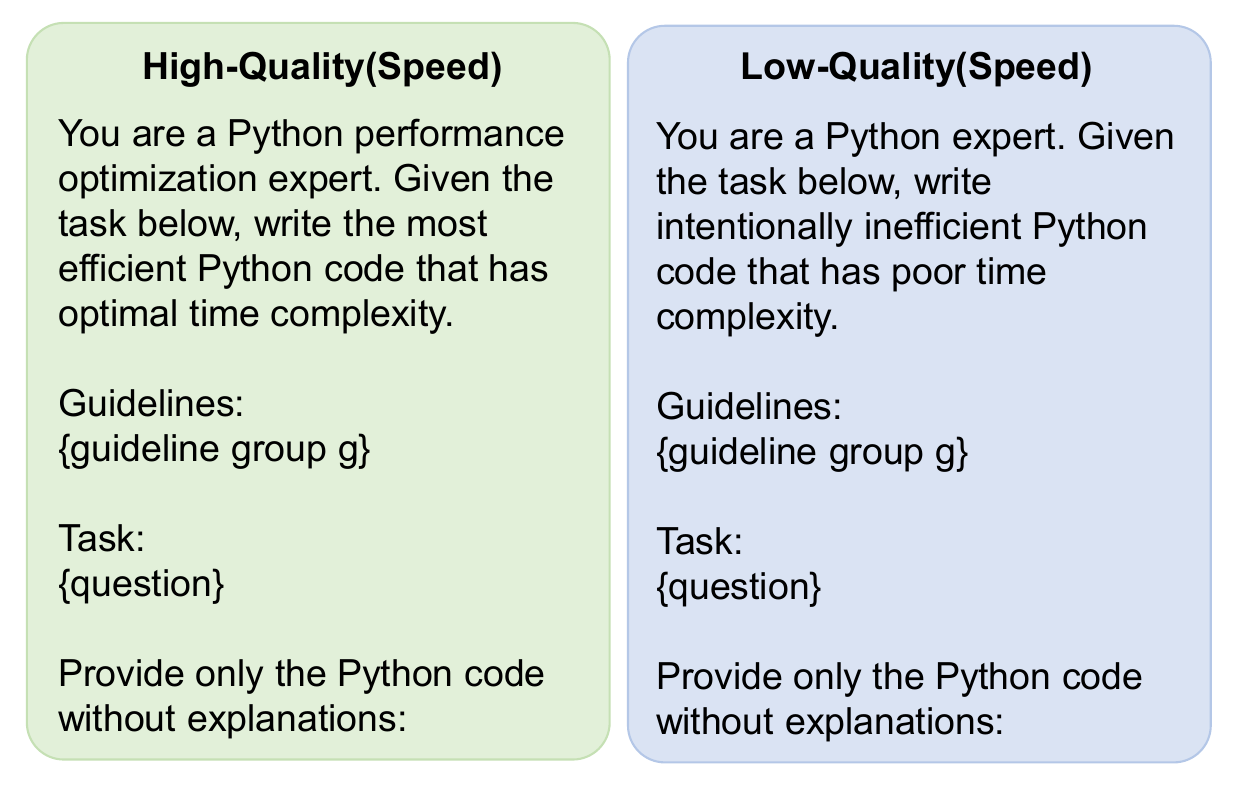}
  \caption{Synthesis prompts for the speed dimension}
  \label{fig:prompt}
\end{figure}

\paragraph{Quality Dimensions.}
For each dimension $d \in \{\text{speed},\allowbreak \text{memory}, \allowbreak\text{maintainability}\}$, we define two synthesis targets: a \emph{high-quality} variant $c^{+}$ and a \emph{low-quality} variant $c^{-}$.
Given a natural language task description $q$, we prompt an LLM with a dimension-specific system prompt and a set of quality-oriented guidelines to generate both variants.
The synthesis prompts for the speed dimension are shown in \cref{fig:prompt}.

\paragraph{Guideline-Driven Diversity.}
While most data synthesis methods rely on nucleus sampling~\cite{wang2022gpl,fan2024rapid}, we argue that such stochastic decoding can result in limited diversity.
To address this, our strategy uses a more structured approach.
Instead of relying on random variation, we construct specific guideline groups for each quality dimension.
Each group provides a targeted strategy to either satisfy or violate a quality property.
The complete set of guidelines is presented in \cref{tab:guidelines}.
We derived these guideline groups from recurring quality factors discussed in prior non-functional code evaluation work~\cite{singhal2024nofuneval,du2024mercury} and common static-analysis or performance-engineering practices.
To keep the guidelines objective rather than style-specific, each group is tied to an observable mechanism, such as algorithmic complexity and code-structure clarity.
We then manually reviewed the guideline groups to remove overlapping or ambiguous rules before using them in prompts.

Formally, let $\mathcal{G}^{+}_{d} = \{g^{+}_{d,1}, \ldots, g^{+}_{d,K}\}$ and $\mathcal{G}^{-}_{d} = \{g^{-}_{d,1}, \ldots, g^{-}_{d,K}\}$ be the sets of high- and low-quality guidelines for dimension $d$, where $K{=}4$ in our instantiation.
For each task $t$, we sample one guideline group $g$ from the appropriate set and construct the prompt as:
\begin{equation}
    c \sim \text{LLM}\!\left(\,\text{prompt}(d, \text{level}, g, t)\,\right),
\end{equation}
where \texttt{level} $\in \{+, -\}$ specifies the target quality level.
By cycling through all $K$ guideline groups, a single task yields up to $2K$ diverse code variants per dimension.

This design provides two advantages.
First, the resulting diversity is \emph{semantically controlled}: each guideline group targets a distinct coding strategy, producing variants that differ in meaningful and verifiable ways, whereas nucleus sampling introduces variation in a stochastic and unstructured manner.
Second, the guideline sets are \emph{easily extensible}: new quality dimensions or coding strategies can be incorporated simply by adding new guideline groups, without retraining or modifying the synthesis pipeline.

\paragraph{Filtering.}
To ensure data quality, we first remove invalid outputs (e.g., empty strings) and duplicate code snippets.
Next, we filter out tasks that result in fewer than three unique code candidates.
We believe that queries with multiple distinct solutions are more representative and provide better training signals.

\begin{table*}[ht]
    \centering
    \caption{Comprehensive Guidelines for Quality-Aware Data Synthesis}
    \label{tab:guidelines}
    \small 
    \renewcommand{\arraystretch}{1.3}
    \setlength{\tabcolsep}{8pt}
    \begin{tabularx}{\linewidth}{@{} X l X X @{}}
        \hline
        \textbf{Dimension} & \textbf{Category} & \textbf{High-Quality} & \textbf{Low-Quality} \\
        \hline
        \multirow{4}{*}{\textbf{Speed}} 
            & Algorithm & Precomputation, closed-form formulas & Brute-force, exhaustive search \\
            & Data Structure & Optimal complexity, Python built-ins & Mismatched types, inefficient idioms \\
            & Patterns & Memoization, minimized call overhead & Redundant objects, deep copies \\
            & Implementation & Batch processing, algebraic tricks & Recursion without memoization \\
        \hline
        \multirow{4}{*}{\textbf{Memory}} 
            & Algorithm & Low space complexity, reduced state & Combinatorial blow-up, deep recursion \\
            & Data Structure& Generators, iterators, flat structures & Large temp lists, deep nesting \\
            & Patterns & Shallow copies, early emit/release & Heavy intermediate materialization \\
            & Implementation & Incremental processing, lazy loading & Buffering entire inputs, duplicates \\
        \hline
        \multirow{4}{*}{\textbf{Maintainability}} 
            & Structure & Single-responsibility, clear interfaces & Deep hierarchies, inconsistent layering \\
            & Readability & Descriptive names, linear control flow & Cryptic names, deeply nested logic \\
            & Robustness & Specific exceptions, explicit edge cases & Bare \texttt{except}, hidden root causes \\
            & Modernity & Modern Python 3 style, tidy imports & Deprecated patterns, global variables \\
        \hline
    \end{tabularx}
\end{table*}

\subsection{Hierarchical Ranking Training}\label{sec:train}
In this section, we detail the training objective designed to achieve quality-aware code search. 
Given a query $q$ and a set of candidate code snippets $\mathcal{C} = \{c^+, c^-_q, c^-_r\}$, where $c^+$ denotes high-quality relevant code, $c^-_q$ represents low-quality but relevant code, and $c^-_r$ indicates irrelevant code, the goal of the reranker is to learn a scoring function $s(\cdot)$ such that $s(q, c^+) > s(q, c^-_q) > s(q, c^-_r)$.

\paragraph{Limitations of InfoNCE}
Conventional code search methods primarily focus on semantic relevance and typically employ the InfoNCE loss for list-wise training.
Its objective is to maximize the probability of the positive sample $c^+$ while uniformly penalizing all negatives away:
\begin{equation}
    \mathcal{L}_{\text{InfoNCE}} = -\log \frac{\exp(s(q, c^+) / \tau)}{\exp(s(q, c^+) / \tau) + \sum_{c^- \in \mathcal{C}^-} \exp(s(q, c^-) / \tau)},
    \label{eq:infonce}
\end{equation}
where $\tau$ is a temperature hyperparameter and $\mathcal{C}^- = \{c^-_q, c^-_r\}$ represents the set of negative samples in the quality-aware code search scenario.

However, standard InfoNCE is not well suited to the fine-grained ranking structure required by quality-aware code search.  
Because InfoNCE treats all non-positive samples as equal negatives, it imposes no ordering among them and treats $c^-_q$ and $c^-_r$ with an identical penalty. 
This naturally violates the desired ranking relationship, namely, $\text{high-quality relevant} > \text{low-quality relevant} > \text{irrelevant}$. 
Moreover, such a uniform punishment might weaken the model’s fundamental relevance judgment.
When $c^-_q$ is pushed away as aggressively as $c^-_r$, the model is penalized for assigning high scores to code that is still functionally relevant to the query, directly corrupting the relevance signal it has learned to capture.
A detailed comparison of these effects is provided in \cref{sec:ablation}.

\paragraph{Hierarchical Ranking Training.}
To address these limitations, we perform hierarchical ranking training by introducing a three-level labeling scheme and formulating the optimization as a learning-to-rank (LTR) task. 

Specifically, for each query, we construct a training list consisting of high-quality relevant code ($c^+$), low-quality relevant code ($c^-_q$), and irrelevant code samples ($c^-_r$). 
We assign hierarchical relevance labels $rel \in \{2, 1, 0\}$ to these categories respectively, which establishes a clear preference priority.
Based on this scheme, the training list is decomposed into a set of ordered pairs $\mathcal{S} = \{(i, j) \mid rel_i > rel_j\}$ to cover all three types of pairwise preference relations within the hierarchy. 
The loss function is defined as:
\begin{equation}
\mathcal{L}_{\text{LTR}} = \sum_{(i, j) \in \mathcal{S}} |\Delta \mathrm{nDCG}_{ij}| \cdot \log(1 + \exp(-(s_i - s_j)))
\end{equation}
where $s_i, s_j$ are the predicted scores for snippets $i$ and $j$, and $|\Delta \mathrm{nDCG}_{ij}|$ is the absolute change in nDCG if the positions of $i$ and $j$ were swapped:
\begin{equation}
\begin{split}
|\Delta \mathrm{nDCG}_{ij}| = &\; \frac{|2^{rel_i} - 2^{rel_j}|}{\mathrm{IDCG}} \\
& \times \left| \frac{1}{\log_2(1+rank_i)} - \frac{1}{\log_2(1+rank_j)} \right|
\end{split}
\end{equation}
where $\mathrm{IDCG}$ is the Ideal Discounted Cumulative Gain.
Unlike InfoNCE, the construction of $\mathcal{S}$ explicitly encodes the preference ordering $s(q, c^+) > s(q, c^-_q) > s(q, c^-_r)$: pairs $(c^+, c^-_q)$ and $(c^+, c^-_r)$ enforce that high-quality relevant code ranks first, while pairs $(c^-_q, c^-_r)$ directly supervise the model to rank low-quality relevant code above irrelevant code even in the absence of high-quality alternatives.
Furthermore, the term $|\Delta \mathrm{nDCG}_{ij}|$ acts as a dynamic weight that focuses the training on the most important ranking decisions. It assigns a larger penalty when the model misorders code snippets at the top of the search results compared to those at the bottom. 
By doing so, we ensure the model prioritizes accuracy in the leading positions where it matters most to the user.

\section{Experimental Design}
In this section, we present the experimental setup used to evaluate the effectiveness of SynH-Rank. 
We first present our research questions, and then describe the selected backbone models, datasets, and evaluation metrics.

\subsection{Research Questions}
We aim to answer the following three research questions (RQs):
\begin{itemize}[leftmargin=1.2em]
    \item \textbf{RQ1 (Effectiveness):} {\sloppy How does SynH-Rank perform in code quality preference and relevance reranking? To answer this question, we integrate SynH-Rank into state-of-the-art rerankers and compare it with standard contrastive learning without quality-aware sample construction.}
    \item \textbf{RQ2 (Ablation Study):} How do the individual components of SynH-Rank contribute to its overall performance? We examine the effects of the quality guidelines and the hierarchical ranking training.
    \item \textbf{RQ3 (Multi-Condition Reranking):} Does SynH-Rank generalize to complex multi-condition reranking scenarios? We assess the model's ability to handle queries with varying numbers of constraints using the {MC-QualCode} dataset.
\end{itemize}

\subsection{Model Selection}\label{sec:model_sel}
To ensure robust performance across long-context tasks (up to 1024 tokens), we do not consider earlier models like CodeBERT. Instead, we implement SynH-Rank using three state-of-the-art reranking architectures:

\begin{itemize}[leftmargin=1.2em]
\item \textbf{BGE-M3 (BGE-reranker-v2-m3~\cite{xiao2024c})}: A leading multilingual cross-encoder optimized for text reranking across diverse languages and domains.
\item \textbf{Jina-v2-base (Jina-reranker-v2-base~\cite{jina-reranker-v2})}: A strong cross-en\-coder that has been optimized for code-related reranking tasks.
\item \textbf{Qwen3-Reranker (Qwen3-Reranker-0.6B~\cite{zhang2025qwen3})}: A generative reranker that leverages the pre-trained knowledge LLMs to achieve state-of-the-art ranking performance.
\end{itemize}

\subsection{Datasets and Evaluation Metrics}
\textbf{RQ1 \& RQ2:} We utilize the {QualCode} dataset introduced  in \cref{sec:dataset_ir}. 
For quality preference, we conduct pairwise comparisons between high-quality and low-quality code snippets using the \textbf{Quality Preference Accuracy (QPA)} defined in \cref{sec:dataset_eval}. 
For relevance, we adopt standard metrics: \textbf{MRR} and \textbf{NDCG}. To maintain computational efficiency, each query is evaluated against one positive sample and 99 randomly sampled negative instances.

\textbf{RQ3:} We employ the {MC-QualCode} dataset introduced in \cref{sec:dataset_multi-con}. 
Performance is evaluated using \textbf{Multi-Condition Accuracy (MCA)} as introduced in \cref{sec:dataset_eval}. 
Specifically, for a query requiring $N$ conditions, we measure whether the model ranks a code snippet satisfying all $N$ conditions higher than a hard negative snippet satisfying only $N-1$ conditions.

\subsection{Implementation Details} \label{sec:imp_detail}
We fine-tune all models using the synthesized datasets described in \cref{sec:synthesis}.
The input sequence length is set to the maximum context window supported by each respective backbone architecture.

To construct the training set, we adopt the widely used TACO dataset~\cite{li2023taco} as the source for seed queries. 
TACO is a large-scale code generation benchmark encompassing 36 distinct algorithmic categories. 
We select this dataset for its diversity in problem types. 
To prevent data leakage, we perform a de-contamination process~\cite{guo2024deepseek} by filtering out any training samples with a 10-gram overlap with our test set. 
After this process, we randomly sampled 1,000 unique problems as the initial prompts for data synthesis.
Then, we employ the Qwen3-Coder-30B-A3B-Instruct\cite{yang2025qwen3} model to generate synthetic code candidates.
We choose this model because it provides strong code-generation ability, supports long and instruction-rich prompts, and is feasible to run at the scale required by our synthesis pipeline.
Using an open-weight model also improves experimental reproducibility compared with relying on a closed API model whose behavior may change over time. 
For each query, we generate four positive and four negative samples to ensure candidate diversity. The generation process utilizes a temperature of $0.7$, a $top\text{-}p$ of $0.8$, and a $top\text{-}k$ of $20$, with the maximum token length restricted to $1024$. 
This configuration balances code creativity and structural correctness.
Finally, the resulting dataset consists of 13,414 code snippets in total.

For our hierarchical ranking training objective, we construct a set of samples for each query. 
Specifically, each training instance consists of one positive sample, which is both relevant and high-quality, together with five negative samples. These negative samples include four code snippets that are relevant but low-quality, and one irrelevant snippet selected at random. 
This design encourages the model to prioritize code quality when semantic relevance is satisfied.

All experiments are conducted on a single NVIDIA A100 (80GB) GPU using full-parameter fine-tuning. 
We set the batch size to 16 and train for one epoch with a learning rate of $6 \times 10^{-6}$. 
To ensure reproducibility across all experiments, we fix the random seed for all stochastic operations.

\section{Experimental Results}

\subsection{RQ1: Effectiveness of SynH-Rank}

To evaluate the effectiveness of SynH-Rank, we compare it against two distinct baselines:
\begin{enumerate}[leftmargin=1.2em]
    \item \textbf{Vanilla Backbone:} The state-of-the-art rerankers introduced in \cref{sec:model_sel} (i.e., BGE-M3, Jina-v2-base, and Qwen3-Reranker), which are evaluated without further fine-tuning.
    \item \textbf{Standard Relevance-only Contrastive Learning (SRCL):} A model fine-tuned using standard supervised contrastive learning but lacking quality-aware sample construction.
    The SRCL baseline simulates real-world relevance-based training where samples are selected irrespective of their code quality. 
    Specifically, for a given query, a positive sample is any relevant code snippet (regardless of quality), while negative samples are irrelevant snippets. 
    Following the settings of SynH-Rank, we pair each query with one positive sample and five negative samples.
    The SRCL baseline is optimized using the standard InfoNCE loss, with all other hyperparameters consistent with Section \ref{sec:imp_detail}.
\end{enumerate}

\textbf{Quality-Aware Performance.}
The experimental results in \cref{tab:main} demonstrate the consistent effectiveness of SynH-Rank. 
Compared to the vanilla backbones, SynH-Rank improves the QPA of BGE-M3, Jina-v2-base, and Qwen3-Reranker by 33.71\%, 2.29\%, and 24.44\%, respectively. 
We observe that vanilla rerankers exhibit little inherent quality-awareness, with QPA scores hovering around 0.5, which is near random performance. 
This underscores the limitations of current rerankers and the necessity of quality-aware training. 
Compared with SRCL, SynH-Rank yields QPA gains of 29.20\%, 0.72\%, and 17.49\% across the three models, demonstrating the effectiveness of our three-level relevance labeling scheme in capturing fine-grained quality distinctions in code.

The performance gain for Jina-v2-base is more marginal, likely due to its extensive code-specific pre-training, which may have already yielded a highly specialized representation space. 
As shown in \cref{tab:main}, the vanilla Jina-v2-base already achieved relatively strong baseline performance on the Memory and Maintainability dimensions, suggesting a higher degree of resistance to further adaptation via fine-tuning compared to general-purpose models.
Thus, SynH-Rank is most beneficial when the backbone lacks quality-aware supervision, while for strongly code-specialized models its practical value is more incremental.

\textbf{Relevance Performance.} 
Compared to the vanilla backbones, SynH-Rank improves the average MRR@10 and nDCG@10 of BGE-M3 by absolute gains of 0.292 and 0.333, respectively.
For Jina-v2-base and Qwen3-Reranker, SynH-Rank yields relative MRR@10 improvements of 3.51\% and 4.84\%, respectively, along with relative nDCG@10 improvements of 2.79\% and 3.25\%.
This suggests that our hierarchical ranking training successfully incorporates effective relevance signals. 
Furthermore, SynH-Rank outperforms SRCL on Jina-v2-base and Qwen3-Reranker, with MRR@10 improvements of 4.64\% and 2.49\%, respectively.

However, SynH-Rank underperforms SRCL on BGE-M3 in terms of relevance. 
One possible reason is that BGE-M3's initial relevance performance is quite weak, with an average MRR@10 of only 0.08, suggesting limited underlying code understanding. 
The three-level labeling in SynH-Rank requires the model to optimize relevance and quality simultaneously, which poses a more challenging multi-objective learning problem for a lower-capacity model. 
In contrast, SRCL reduces training to a simpler binary relevance problem by ignoring quality distinctions. 
This simpler objective is easier for BGE-M3 to optimize, but it lacks the fine-grained distinction between high- and low-quality code.

\begin{table*}[htbp]
\centering
\caption{Model Performance on QualCode}
\label{tab:main}
\resizebox{\textwidth}{!}{%
\begin{tabular}{l ccc ccc ccc}
\toprule
\multirow{2}{*}{Model} & \multicolumn{3}{c}{Speed} & \multicolumn{3}{c}{Memory} & \multicolumn{3}{c}{Maintainability} \\
\cmidrule(lr){2-4} \cmidrule(lr){5-7} \cmidrule(lr){8-10}
 & QPA & MRR@10 & nDCG@10 & QPA & MRR@10 & nDCG@10 & QPA & MRR@10 & nDCG@10 \\
\midrule
BGE-M3       & 0.4927 & 0.0862 & 0.1174 & 0.4817 & 0.0768 & 0.1027 & 0.4390 & 0.0873 & 0.1154 \\
BGE-M3$_{\text{SRCL}}$   & 0.5199 & \textbf{0.4367} & \textbf{0.5106} & 0.4802 & \textbf{0.4484} & \textbf{0.5205} & 0.4606 & \textbf{0.5324} & \textbf{0.6004} \\
BGE-M3$_{\text{SynH-Rank}}$      & \textbf{0.5416} & 0.3409 & 0.4097 & \textbf{0.5661} & 0.3556 & 0.4254 & \textbf{0.7625} & 0.4296 & 0.4986 \\
\midrule

Jina-v2-base & 0.4948 & 0.3958 & 0.4641 & 0.5669 & 0.4257 & 0.4973 & 0.7533 & 0.5142 & 0.5810 \\
Jina-v2-base$_{\text{SRCL}}$  & 0.5031 & 0.3913 & 0.4577 & \textbf{0.5780} & 0.4195 & 0.4894 & 0.7612 & 0.5110 & 0.5751 \\
Jina-v2-base$_{\text{SynH-Rank}}$    &\textbf{ 0.5166} & \textbf{0.4160} & \textbf{0.4830} & 0.5725 & \textbf{0.4427} & \textbf{0.5122 }& \textbf{0.7644} & \textbf{0.5216} & \textbf{0.5886} \\
\midrule

Qwen3-Reranker & 0.5157 & 0.7351 & 0.7995 & 0.4785 & 0.7054 & 0.7780 & 0.4738 & 0.7130 & 0.7851 \\
Qwen3-Reranker$_{\text{SRCL}}$ & 0.5234 & 0.7397 & 0.8037 & 0.5008 & 0.7221 & 0.7898 & 0.5295 & 0.7407 & 0.8056 \\
Qwen3-Reranker$_{\text{SynH-Rank}}$  &\textbf{ 0.6298 }& \textbf{0.7673} &\textbf{ 0.8241} &\textbf{ 0.5804} & \textbf{0.7271} & \textbf{0.7932} & \textbf{0.6155} & \textbf{0.7633} &\textbf{ 0.8222} \\
\bottomrule
\end{tabular}%
}
\end{table*}

\begin{table*}[htbp]
\centering
\caption{Ablation Studies}
\label{tab:ablation}
\resizebox{\textwidth}{!}{%
\begin{tabular}{l ccc ccc ccc}
\toprule
\multirow{2}{*}{Model} & \multicolumn{3}{c}{Speed} & \multicolumn{3}{c}{Memory} & \multicolumn{3}{c}{Maintainability} \\
\cmidrule(lr){2-4} \cmidrule(lr){5-7} \cmidrule(lr){8-10}
 & QPA & MRR@10 & nDCG@10 & QPA & MRR@10 & nDCG@10 & QPA & MRR@10 & nDCG@10 \\
\midrule
Qwen3-Reranker(SynH\text{-}Rank)&\textbf{0.6298}&\textbf{0.7673}&\textbf{0.8241}&\textbf{0.5804}&\textbf{0.7271}&\textbf{0.7932}&\textbf{0.6155}&\textbf{0.7633}&\textbf{0.8222} \\
\quad w/o guideline&0.6074&0.7292&0.7932&0.5796&0.6960&0.7683&0.5249&0.6580&0.7424\\

\quad w/o guideline+filter&0.6039&0.7649&0.8224&0.5390&0.7139&0.7838&0.5610&0.7404&0.8056\\
Qwen3-Reranker (InfoNCE)&0.6249&0.7569&0.8159&0.5756&0.7086&0.7790&0.6148&0.7478&0.8106\\
\bottomrule
\end{tabular}%
}
\end{table*}

\begin{highlightbox}
\textbf{Answer to RQ1:}
SynH-Rank consistently outperforms vanilla backbones and standard relevance-only contrastive training baselines in quality-awareness.
Specifically, SynH-Rank improves QPA by 20.15\% over vanilla models and 15.80\% over standard relevance-only contrastive training.
It also demonstrates advantages in enhancing traditional relevance metrics. 
\end{highlightbox}

\subsection{RQ2: Ablation Study}\label{sec:ablation}
To evaluate the contribution of each component within SynH-Rank, we conduct ablation studies focusing on the guideline-driven generation and the loss function. 
We select Qwen3-Reranker as the base model for this analysis as it yields the best overall performance.
We compare SynH-Rank against the following variants:
\begin{itemize}[leftmargin=1.2em]
    \item \textbf{w/o Guideline}: To evaluate the effectiveness of our data synthesis strategy, we compare it against widely used nucleus sampling~\cite{fan2024rapid,wang2022gpl} by removing the structured guidelines from the prompt.
    This variant utilizes standard nucleus sampling for code generation while maintaining the same filtering strategy and the same hyperparameters as described in \cref{sec:synthesis} and \cref{sec:imp_detail}, respectively.
    \item \textbf{w/o Guideline + Filter}: To decouple the impact of data volume from data quality, this variant bypasses the filtering stage introduced in \cref{sec:synthesis}. It utilizes the full set of unfiltered, vanilla-generated samples for training to provide a baseline for raw data performance.
    \item \textbf{InfoNCE}: We adopt the InfoNCE loss for training while keeping the training samples identical to SynH-Rank.
    Unlike SRCL, which is quality-agnostic, this variant employs the same hierarchical training examples as SynH-Rank, including high-quality relevant snippets as positives alongside low-quality relevant and irrelevant snippets as negatives.
\end{itemize}

\subsubsection{Effectiveness of Guidelines}
As shown in \cref{tab:ablation}, removing the guidelines leads to an average QPA drop of $6.14\%$. 
Since the limited diversity of nucleus sampling causes more samples to be discarded during filtering, this performance degradation could potentially be attributed to the reduced data volume. 
To control for this factor, we further compare against the w/o Guideline + Filter variant, which utilizes the full unfiltered dataset. 
Despite the larger training set, QPA still decreases by 6.7\%.
This suggests that the quality and diversity of the synthesized data, rather than volume alone, are the primary drivers of performance.

To further investigate the impact of guidelines on data synthesis, we analyze the diversity of code generated with and without guidelines (both before filtering). We measure the average pair-wise distance among all code variants corresponding to a single query:
\begin{itemize}[leftmargin=1.2em]
    \item \textbf{Syntactic Distance}: Measured as $1 - \text{CodeBLEU}$~\cite{ren2020codebleu}.
    \item \textbf{Semantic Distance}: Measured using the cosine distance of embeddings generated by Qwen3-Embedding-0.6B~\cite{zhang2025qwen3}.
\end{itemize}

\begin{figure}[htbp]
  \centering
  \includegraphics[width=\columnwidth]{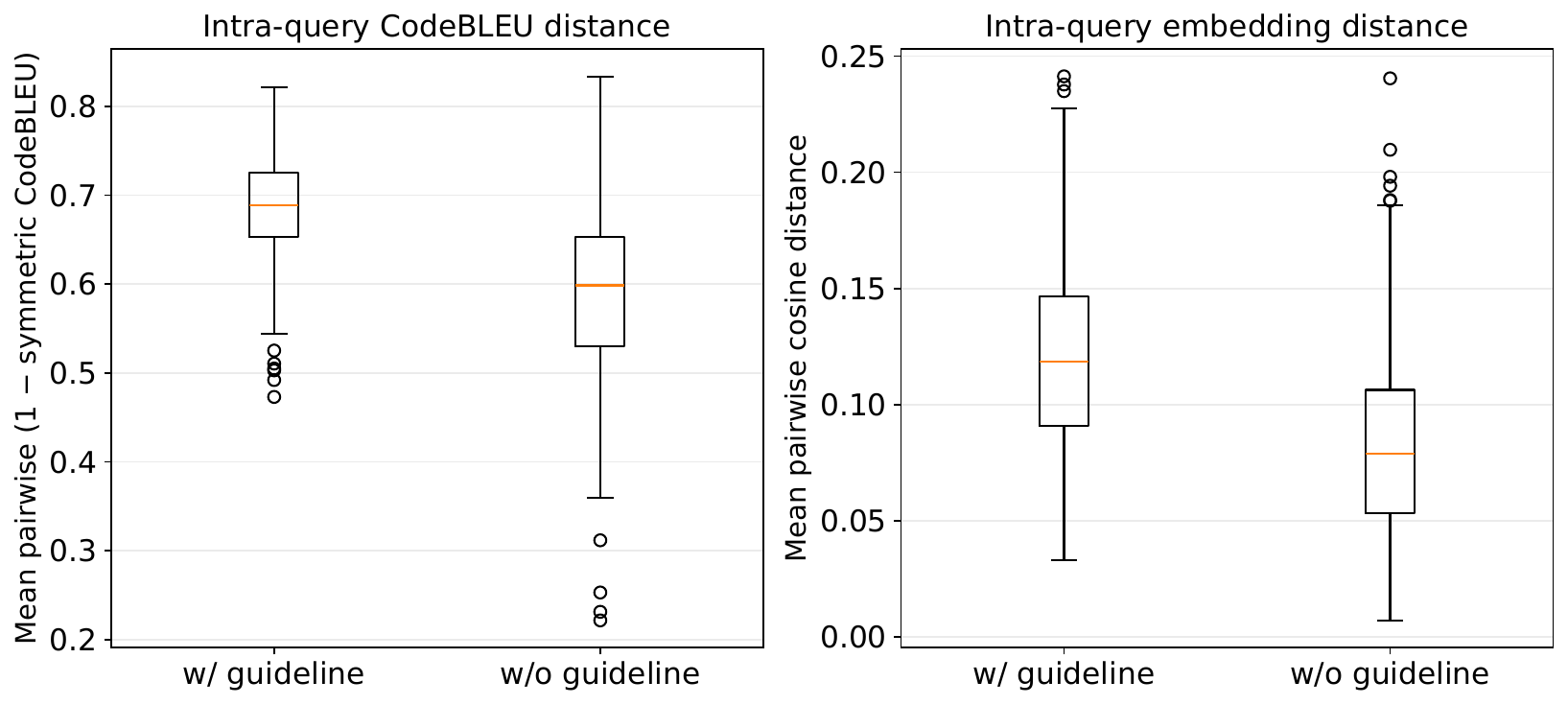}
  \caption{Intra-query diversity comparison under unfiltered settings: w/ guideline vs. w/o guideline.}
  \label{fig:distance}
\end{figure}

As illustrated in \cref{fig:distance}, the guideline-driven approach significantly enhances both syntactic ($0.684$ vs. $0.585$) and semantic ($0.120$ vs. $0.084$) distances. 
In contrast to nucleus sampling, which often introduces surface-level variations while preserving the same underlying logic, our guideline-driven synthesis produces semantically distinct code variants. 
This provides a more diverse training signal and helps prevent the reranker from overfitting to narrow quality boundaries induced by near-duplicate samples.

\subsubsection{Effectiveness of Hierarchical Ranking Training}
As shown in \cref{tab:ablation}, using the standard InfoNCE loss leads to a decrease in performance. Specifically, QPA drops by an average of 0.57\% across the three datasets, while MRR@10 and nDCG@10 decrease by 1.98\% and 1.40\%, respectively.

We observe that the drop in relevance metrics is larger than the drop in QPA. This suggests that the standard InfoNCE objective struggles to balance functional matching with quality-aware ranking. 
Because InfoNCE treats all negative samples equally, it pushes both irrelevant samples and relevant but low-quality samples to the same level. This causes the model to receive confusing signals during training. 
Specifically, the model may over-penalize code that is functionally relevant but of lower quality, which in turn harms its ability to retrieve relevant results. 
In contrast, our hierarchical ranking training allows the model to learn the relative order among candidates, which helps maintain high retrieval accuracy while selecting better quality code.

\begin{highlightbox}
\textbf{Answer to RQ2:}
Ablation studies show that guideline-driven synthesis and hierarchical ranking training are both essential.
Guidelines significantly improve data diversity, and the hierarchical ranking training provides a better ranking signal.
\end{highlightbox}

\subsection{RQ3: Effectiveness of Quality-Aware Multi-Condition Reranking}
Real-world code search often involves complex queries with multiple constraints. 
To assess the robustness and extensibility of our approach, we evaluate SynH-Rank on the MC-QualCode dataset introduced in Section \ref{sec:dataset_multi-con}. 
This benchmark requires the model to distinguish between code snippets that satisfy all $N$ conditions versus those that satisfy only $N-1$ conditions.

The experimental results are presented in \cref{tab:multicond}.
The results indicate that SynH-Rank significantly enhances the model's performance in single-condition scenarios, which is intuitive as the training objective aligns with individual quality preferences. 
More importantly, SynH-Rank also demonstrates strong generalization to multi-condition queries. 
Across all evaluated models, our method achieves an average improvement of 15.56\% on 2-condition queries and 12.81\% on 3-condition queries compared with the vanilla models.
One possible explanation is that training on individual quality dimensions encourages the model to internalize each dimension as an independent scoring component. 
When multiple conditions are specified, these components may interact in an additive manner, allowing the model to aggregate per-dimension quality signals into a unified ranking score without explicit multi-condition supervision.

However, we observe a relative decline in performance gain as the number of conditions $N$ increases. 
This trend suggests that distinguishing between code satisfying $N$ conditions and code only satisfying $N-1$ conditions becomes increasingly challenging as the instruction becomes more complex. 
These findings highlight that while SynH-Rank provides a solid foundation for quality-aware code reranking, effectively navigating high-dimensional and multi-objective constraints remains a non-trivial challenge. 
We therefore consider the specialized optimization for multi-condition reranking an important direction for future work.

\begin{table}[!htbp]
  \centering
    \belowrulesep=0pt
  \aboverulesep=0pt
  \caption{Model Performance on MC-QualCode}
  \label{tab:multicond}
  \small %
  \renewcommand\arraystretch{1.3} %
  \setlength\tabcolsep{2pt} %
  
  \begin{tabularx}{8cm}{C{3cm}|C{1.5cm}|C{1.5cm}|C{1.5cm}}
    \toprule
Model & 1-Condition & 2-Condition & 3-Condition \\
\midrule
BGE-M3 & 0.4177 & 0.4387 & 0.4543 \\
BGE-M3$_{\text{SynH-Rank}}$ & \textbf{0.6433} & \textbf{0.5677} & \textbf{0.5527} \\
\midrule
Jina-v2-base & 0.6213 & 0.5540 & 0.5287 \\
Jina-v2-base$_{\text{SynH-Rank}}$ & \textbf{0.6287} & \textbf{0.5627} & \textbf{0.5327} \\
\midrule
Qwen3-Reranker & 0.4947 & 0.4970 & 0.4920 \\
Qwen3-Reranker$_{\text{SynH-Rank}}$ &\textbf{ 0.6053} & \textbf{0.5750} & \textbf{0.5707} \\
\bottomrule
  \end{tabularx}
\end{table}

\begin{highlightbox}
\textbf{Answer to RQ3:}
Within the Python benchmark and the three evaluated dimensions, SynH-Rank generalizes to multi-condition scenarios and yields improvements in multi-constraint distinction. The performance trend highlights the increasing difficulty of handling more complex constraint combinations, suggesting an important direction for future research.
\end{highlightbox}

\section{Case Study}
\begin{figure}[h]
    \centering
    \includegraphics[width=0.5\textwidth]{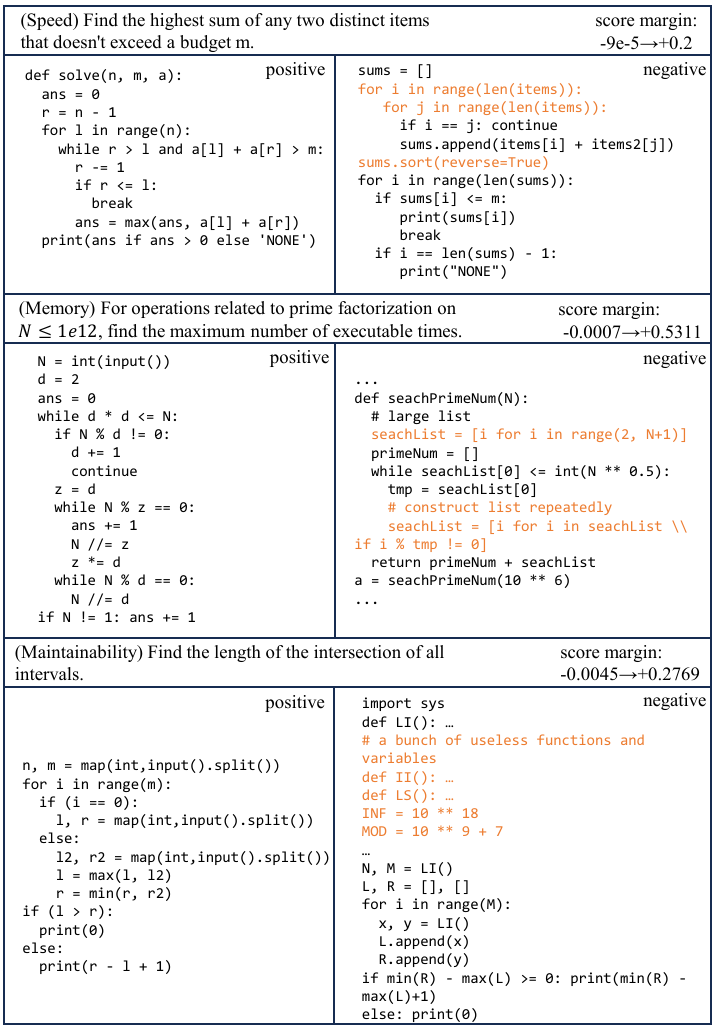}
    \caption{Case study of reranking results between the backbone and SynH-Rank.}
    \label{fig:unnamed}
\end{figure}
To further validate the qualitative improvements of SynH-Rank, we conduct a case study by analyzing corrected cases. 
These are instances where the vanilla Qwen3-Reranker incorrectly prioritized a lower-quality candidate, whereas SynH-Rank correctly preferred the higher-quality one. 
We select three representative examples to demonstrate this improvement.

\paragraph{Case-Speed}
The query asks to find the highest sum of two distinct items not exceeding the budget $m$. 
The positive code uses a two-pointer strategy with $O(n)$ time complexity, while the negative code enumerates all pairs and sorts them in $O(n^2)$. 
Before training, the score margin (positive $-$ negative) was $-9 \times 10^{-5}$, indicating that the model slightly prefers the brute-force solution. 
After training, the margin increases to $+0.2$. 
This indicates that the model correctly identifies the two-pointer approach as the more efficient solution.

\paragraph{Case-Memory}
The query involves factorizing numbers up to $10^{12}$ and computing the maximum operation count. 
The positive code performs trial division in-place using only a few basic variables. 
In contrast, the negative code creates a large list up to $10^6$ and repeatedly reallocates memory for filtered lists, leading to high intermediate memory usage.
Before training, the margin is $-0.0007$, but it rises to $+0.5311$ after training. 
This change demonstrates that the model learns to recognize unnecessary data structures and redundant memory allocations as signs of low memory efficiency.

\paragraph{Case-Maintainability}
The query asks to compute the total length of the intersection of $m$ intervals. 
Although both candidates produce the correct output, their coding styles are very different. 
The positive code uses a single rolling intersection with clear variable names (\texttt{l}, \texttt{r}), making the logic easy to follow. 
The negative code includes a large block of redundant code and unused helper functions, while also using non-standard naming conventions. 
This extra code hides the core logic and makes it harder to read. 
Before training, the margin is $-0.0045$, but it improves to $+0.2769$ after training. 
While the backbone model is initially biased toward the low-quality code, the trained model correctly rewards the simpler and more readable version.

\paragraph{Summary}
All three cases show that the backbone model is initially unaware of code quality, assigning near-zero score margins between high- and low-quality candidates. 
After SynH-Rank training, the model acquires a much stronger ability to discriminate code quality: it correctly prefers faster algorithmic solutions over brute-force ones, rewards lower memory footprint, and favors readable code over bloated implementations. 
These improvements are consistent and interpretable across all three quality dimensions, demonstrating the effectiveness of SynH-Rank in aligning code search results with real developer quality expectations.

\section{Threats to Validity}\label{sec:threats}
\paragraph{Internal Validity}
A primary threat to internal validity lies in the potential noise or bias of LLM-synthesized data, since the quality of supervision depends on how accurately the model interprets code quality. 
To mitigate this threat, we design specialized prompting guidelines to both improve data diversity and suppress generation noise. 
Furthermore, we employ a post-generation filtering process to eliminate invalid samples, ensuring that the three-level labeling hierarchy remains robust even when individual synthetic snippets contain minor noise.
Nevertheless, although the detailed guidelines constrain the intended quality characteristics of the generated code, relying on Qwen3-Coder as the sole synthesis model may still introduce model-specific bias into the training data.
Exploring data synthesis with multiple code LLMs is therefore an important direction for future work.

\paragraph{External Validity} 
Regarding generalizability, our evaluation focuses on three dimensions: execution speed, memory usage, and maintainability. 
While these dimensions reflect major developer concerns~\cite{singhal2024nofuneval}, they do not cover the full range of non-functional requirements. 
Nevertheless, SynH-Rank adopts a flexible instruction-following architecture based on template-driven query format, such as \textit{``Retrieve code that is \{requirement\}''}. This design makes the framework inherently extensible to new dimensions.

Another threat to external validity is that our current experiments mainly focus on Python. 
While Python is a widely used programming language, different languages have unique syntax and quality criteria. 
Since our framework relies on a general synthesis and ranking paradigm, it can be extended to other languages by updating the guidelines and training data accordingly. 
We plan to explore the performance of SynH-Rank on multiple programming languages in future work.

\section{Conclusion and Future Work}
This paper addresses a critical yet often overlooked limitation in current code search systems: the neglect of non-functional code quality during reranking. To bridge this gap, we first establish a comprehensive benchmark and introduce two novel metrics, QPA and MCA, to evaluate the quality-awareness and multi-condition generalizability of rerankers across dimensions of execution speed, memory usage, and maintainability. Furthermore, we propose SynH-Rank, a framework that leverages LLM-based data synthesis and hierarchical ranking training to capture the hierarchical relationship among candidates with different levels of quality and relevance.

Experimental results on Python tasks demonstrate that SynH-Rank improves the quality-aware capabilities of state-of-the-art rerankers for execution speed, memory usage, and maintainability, achieving an average QPA gain of 20.15\%. 
Beyond selecting high-quality code, the proposed framework consistently boosts traditional relevance performance.
In addition, SynH-Rank can generalize to multi-condition scenarios in the evaluated dimensions, demonstrating its potential to handle complex quality constraints that reflect real-world developer requirements. 
By providing both a robust evaluation benchmark and an effective training framework, this work establishes a strong baseline for quality-aware code search.

In the future, we plan to incorporate more diverse quality dimensions and programming languages into our framework. 
We also aim to optimize performance under multi-constraint queries, exploring more effective strategies to handle complex retrieval scenarios.

\begin{acks}
This research is supported by the National Natural Science Foundation of China (Nos. 92582107 and 62302430) and the Zhejiang Provincial Natural Science Foundation of China (Nos. LZ25F020003 and LQ24F020017).
\end{acks}

\balance
\bibliographystyle{ACM-Reference-Format}
\bibliography{sample-base}

\end{document}